\documentclass[12pt, onecolumn, letterpaper]{article}
\usepackage{mathptmx}        
\usepackage{helvet}          
\usepackage{courier}         
\usepackage{type1cm}         
\usepackage{makeidx}         
\usepackage{graphicx}        
\usepackage{multicol}        
\usepackage[bottom]{footmisc}
\usepackage{url}
\usepackage{cite}
\usepackage{subfigure}
\usepackage{amsmath}
\usepackage{authblk}
\usepackage{geometry}
\usepackage{setspace}
\usepackage{amsfonts} 
\usepackage{mathtools}
\usepackage{lineno}

\makeindex
\title{A complexity perspective on the geographical location of companies: How distance reduce trade between firms}

\author[1,2]{\large Eduardo Viegas}
\author[3,4]{\large Orr Levy}
\author[1,5]{\large Shlomo Havlin}
\author[1,6]{\large Hideki Takayasu}
\author[1]{\large Misako Takayasu}
\affil[1]{\small Department of Mathematical and Computing Science, School of Computing, Tokyo Institute of Technology, Yokohama 226-8502, Japan}
\affil[2]{\small Centre for Complexity Science and Department of Mathematics, Imperial College London, SW7 2AZ, United Kingdom}
\affil[3]{\small HHMI, Yale University School of Medicine, New Haven, CT 06510, USA}
\affil[4]{\small Department of Immunobiology, Yale University School of Medicine, New Haven, CT 06510, USA}
\affil[5]{\small Department of Physics, Bar-Ilan University - Ramat-Gan 5290002, Israel}
\affil[6]{\small Sony Computer Science Laboratories, 3-14-13, Higashi-Gotanda, Shinagawa-ku, Tokyo 141-0022, Japan}

\affil[*]{\small e.viegas11@imperial.ac.uk}

\date{}

\begin{document}

\maketitle

\section*{Abstract}

Geometrical distance is an important constraining factor underpinning the emergence of social and economic interactions of complex systems. 
Yet, agent-based studies supported by granular analysis of distances are limited.
Here, we develop a complexity method that places the real physical world, represented by the actual geographical location of individual firms in Japan, at the epicentre of our research. By combining methods derived from network science (to evaluate the emerging properties of the agents) together with information theory measures (to capture the strength of interaction among these agents), we can systematically analyse a comprehensive dataset of Japanese inter-firm business transactions network and evaluate the effects of spatial features on the structural patterns of the economy.
We find that the normalised probability distributions of distances between interacting firms show a power law like decay concomitant to the sizes of firms and regions, with slower decays in major cities.
Furthermore, small firms would reach large distances to become a customer of large firms while trading between either only small firms, or only large firms, tends to be at smaller distances. However, a time evolution analysis suggests that a level of market optimisation occurs over time as a reduction in the overall average trading distances in last 20 years can be observed. 
Lastly, our analysis concerning the trading dynamics among prefectures indicate that the preference to trade with neighbouring prefectures tends to be more pronounced at rural regions as opposed to the larger central conurbations, leading to the formation of three distinct types of regional geographical clusters.

\section{Introduction}

Studies in both fields of economics and complexity science have provided valuable information on social and economic activities over the past few decades\cite{Boccaletti:2006,Schweitzer:2009}.
In very simplistic terms economics and complexity are complimentary fields. The former tends to be preoccupied with the \emph{behaviour} of agents with regards to the production, consumption and distribution of wealth. In contrast, yet in tandem, the latter tends to be focused on the emerging properties of the \emph{interactions} among the same agents, leading to structures such as small world\cite{Duncan:1998} or scale-free\cite{Barabasi:1999} networks for example.
Such agents, however, are usually located in a real physical world, and therefore, subject to the natural constraints of Earth's geography. Critically, this means that spatial features, such as transportation systems and human communication interaction, could play a notable role in shaping the formation and structure of social and economic interactions. Yet, even with the best infrastructure and with interactions being facilitated by the web (virtual) world, one can hypothesise that some human, social and historical elements associated with neighbourhood and regional cultural may yet still continue to influence on trade.
In this respect, our work is motivated by the fact that a robust, and granular level, trade distance analysis may well provide quantitative insights into the influence of these factors at play.

To date, many studies have focused on the spatial and geographical elements of infrastructure networks - such as subway and logistics - both empirically and analytically\cite{Rozenfeld:2002,Yook:2002,Brunet:2002,Marc:2003,Barrat:2004,Guimera:2005,Jung:2008,Marc:2011}.
However, to our knowledge, little research to date has been specifically designed to quantify the impact of space and geography on the development of real trade inter-firm networks. Instead, studies have been mostly limited to economic models based on aggregated data about properties intrinsic to agents, as opposed to granular level data related to the interaction among these agents. 
A class example of economic analysis is the transaction cost theory, proposed by Ronald H. Coase\cite{Coase:1937}. This is one of the theories that articulates the tendency of firms to gather together in specific areas, such as industrial agglomerations and industrial clusters.
According to this theory, firms tend to locate close to their business partners primarily to reduce costs, not only on those easily measured such as transportation of goods or services, but also on indirect ones such as costs related to human communications to promote projects or make adjustments.
More recently, Nakajima et al.\cite{Nakajima:2012} conducted an empirical study about the localisation of industries in Japan in terms of geometric distances between the whole manufacturing firm using firm-level datasets.
The distance distribution between all pairs of firms on the real network was statistically compared to that of a randomised network, and it was found that the basis of the transaction cost theory of firms did not work for every industry, since other common factors shaping the concentration of industrial activities are interplayed across countries.
The research also confirms that geometric distance is undoubtedly a relevant parameter for firms' business activities.
However, this work did not focus on firms' interactions - such as business transactions - into account nor found any spatial distributions that resemble the form of earlier empirical studies\cite{Yook:2002,Barrat:2004,Guimera:2005,Jung:2008,Marc:2011}.

In a complimentary manner to these studies, our research is not preoccupied with the economic \emph{behaviour} of agents as well as the associated decision making logic and rationale. Instead, we are interested in the questions and phenomena, and we will test in on a large available dataset from Japan in order to identify the core emerging properties of the inter-firm business transaction networks.
Essentially, our work attempts to address three different questions: (a) whether there is a generic functional relationship between geometric distance and frequency of trades (hereinafter, referred to as `trade distance'); (b) whether distance affects trades in a uniform manner across companies of different sizes and industry activities; and (c) whether regional characteristics and patterns influence the dynamics of trade distance.
There are distinct networks that can be constructed and derived from interactions among businesses (i.e. trading, shareholding, competition, etc.). In this research, we centred our work on the inter-firm business transactions layer related to the activities of customers and suppliers of products and services. Within our network (namely, the `Inter-firm Trade Network'), firms are represented by nodes, whereas the resulting money flows arising from the exchange of goods and services between  customers and suppliers form the links (or edges) between these nodes. We note here that the general statistical properties of such networks have been a subject of extensive studies\cite{Takayasu:2007}, but without considering the trade distance dimension.

Below, we divided our study into four distinctively, yet interrelated, analytical methods which are presented in the \textit{Results} section with the relevant and detailed formulae, steps and approach described in the \textit{Methods} section, when necessary. 
Firstly, we present a macro level data analysis by providing the reader with an empirical and analytical description of the interplay and scaling relations between the structure of the Inter-Firm Trade Network, the geometrical distance and the size of firms (or agents), measured by their annual sales. 
We follow with a second sub-section by breaking down the analysis through the prism of industry sectors and prefectures as well as adding the time dimension. 

In the third section we adopt traditional complexity and network randomisation methods\cite{Milo:2003} to highlight and measure the impact of geometrical distance on the distinctive structural features of the Inter-Firm Trade Network from a time and prefecture perspectives. This is done by observing the probability distributions of trade distances, normalised by the randomised network.

Lastly, we use the same data and randomisation methods from the previous section and cross-grain the Inter-Firm Trade Network from company into prefecture level. We then make use of the Mutual Information as a theoretical proxy measure to quantify, distinguish and classify the nature of the relationship among the prefectures\cite{Viegas:2020:Allometric}. This is done in order to test the hypothesis that the distinct economic fundamentals of each prefecture may influence and affect the outcome of geometrical distance trading.


\section{Results}

\subsection{Scaling Relations between Interacting Firms}

\begin{figure}[b!]
  \begin{center}
    \includegraphics[width=1\hsize]{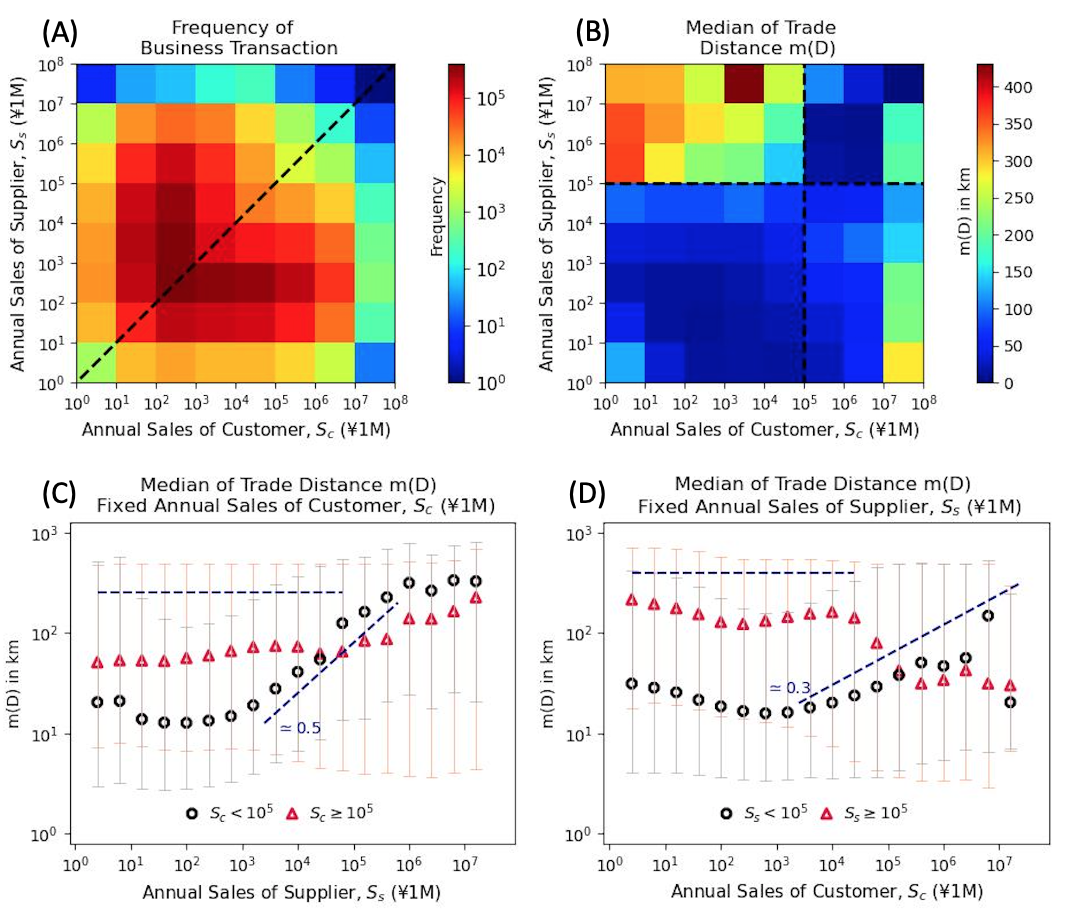} 
    
    \caption{ {\bf The frequency of trade and distances $D$ highly depend on the annual sales of customers $S_{c}$ and annual sales of suppliers $S_{s}$.} \
      ({\bf A}) is a heatmap showing the frequency trades among the $S_{c}$ and $S_{s}$ pairings, where the dotted diagonal line is the symmetry $S_{c}$ = $S_{s}$.
      ({\bf B}) is the equivalent heatmap quantifying the median $m(D)$ of the $S_{c}$ and $S_{s}$ pairing, where the horizontal and vertical lines indicate the break of the $S_{c} \simeq S_{s}$ symmetry.
      ({\bf C}) shows the scaling of the median distance $m(D)$ of firms at different $S_{s}$ levels (above and below the symmetry break) with fixed $S_{c}$, and\
      ({\bf D}) is an equivalent plot with fixed $S_{s}$ and variable $S_{c}$ levels. \
      Each black circle and red triangle represents firms where $S < 10^{5}$ and $S \geq 10^{5}$, respectively.\
      Figures are plotted on a log-log scale for the 2021.\
  \label{fig:1}       }
  \end{center}   
\end{figure}

Scaling relation is a fundamental methodology originating from physics that is commonly used to reveal laws governing functional relationships between distinct, but related, physical quantities. 
Its application tends to lead to a better understanding of systems that have complex structures as well as to a more comprehensible description of their inherent mechanisms. Examples are biological\cite{Brown:2000} and economic\cite{Brock:1999} systems, both of which typically reveal power law scaling behaviours.
Many studies that focused on the characteristics of firms have reported various types of scaling relations\cite{Stanley:1996,Amaral:1997,Takayasu:1998,Axtell:2001,Dongfeng:2005,Watanabe:2013,Takayasu:2014,Goto:2017}.
For example, it was observed that there are scaling laws in firms between the median of annual sales (million yen) $S$, the number of employees $E$, and the number of business transactions $k$. It has been found that $ S \propto E^{1.3},~S \propto k^{1.3} $, and $ E \propto k^{1.0} $\cite{Watanabe:2013}.
Here, we specifically analyse scaling relations between customers and suppliers within the Inter-firm Trade Network in order to better quantify and interpret the relations between geometrical distance and different sizes of firms. Our results are shown in four distinct panels within Fig.~\ref{fig:1}.

Firstly, we compare and contrast the two heatmaps at the top of Fig.~\ref{fig:1} (with methodology details described in subsection~\ref{meth:scaling}). These show ({\bf A}) the frequency of trades and ({\bf B}) the median trade distances $m(D)$ binned by the pairing of the annual sales of customers $S_{c}$ and those of suppliers $S_{s}$ for all existing edges within the Inter-firm Trade Network.
It is noticeable that whereas Panel ({\bf A}) is found to be almost fully symmetrical throughout the boundary $S_{c} = S_{p}$, a similar pattern in Panel ({\bf B}) is only observed below the boundary $S_{c} \simeq S_{p} \simeq 10^{5}$ (represented by the horizontal and vertical dotted lines).
Panel ({\bf A}) also highlights the existence of a relative dominance of middle scaled companies in the overall number of business transactions within the Inter-firm Trade Network.

Following the observed abrupt break in symmetry, we classify and separate the set of edges, or business transactions, into two distinct subsets, where $S_{i} < 10^{5}$ (black circles) and $S_{i} \geq 10^{5}$ (red triangles), for the distinct roles of Suppliers ($S_{i} = S_{s}$) in panel ({\bf C}) and Customers ($S_{i} = S_{c}$) in panel ({\bf D}). 
From these panels, it possible to observe the existence of scaling relationships $D \propto S_{s}^{0.5 \pm 0.1}$ panel ({\bf C}) and $D \propto S_{s}^{0.3 \pm 0.1}$ panel ({\bf D}) for the firms with annual sales below the boundary $S_{i} < 10^5$, and a roughly similar inverse decay for $S_{i} \geq 10^5$.

Here, we emphasise that we make use of the word "association" in a neutral manner so that we do not imply that the individual agents, or firms, are necessarily expressing specific behaviours and preferences as a result of logical decision making processes endogenous to the agents. Instead, we solely focus our attention to the influence of geographical distance to the emerging properties of the Inter-firm Business Network from a structural, and systemic, perspective.

These results seem to indicate that geometric distances tend to play a stronger role in shaping associations (with business partners) on middle sized (or scale) companies if compared to very small or very large companies. By considering the information across all panels, one can also suggest that large customers have a tendency to associate with small suppliers at smaller distances (average 100km), while small customers are usually supplied by firms located at places larger than the average distance (over 200km). In addition, it also appears that the trade distances $D$ between large companies, expressed by upper right of panel ({\bf A}) and a tail of red triangle distributions, are short.  

These structural features can be better understood in the context of the findings described in subsection~\ref{find:mutual}. This is because the core of Tokyo and Osaka have a much smoother decay than those of the other clusters. In addition, larger companies - and related transactions among these companies - are heavily concentrated within the same core clusters (of Tokyo and Osaka). For instance  70\% of large companies (whose sizes are over $S = 10^6$) are located in Tokyo, and they account for over 50\% of the business transactions between these companies. In short, the behaviour of middle scale companies is heavily influenced by the communitarian and midway clusters, where the behaviour of larger companies are mostly shaped by the core clusters.

\subsection{Geometric Proximity between Interacting Firms: Industry Sectors and Prefectures}

Extensive research has benn carried to evaluate industrial localisation, with several methods that have been developed to quantify trade preferences\cite{Hoover:1936,Harris:1954,Holmes:1999}.
However, all these studies have been based on aggregated, coarse grained, data, effectively inhibiting and limiting the analysis of the structural, systemic, features of trade among businesses and firms. In any event, and unsurprisingly, empirical results from these previous studies suggest that the geographical location of firms is undoubtedly a relevant parameter in shaping their business partners trading activities.
So far, however, to our best knowledge, no study has attempted to articulate this phenomenon and quantify it by making use of comprehensive datasets that includes a nationwide business transactions network.
Within such objective in mind, we attempt to observe whether geometric proximity influences in a different manner the generation of business interactions depending on the industry sector. Furthermore, we aim to better understand the potential evolution of trade distances over a relatively long time period (i.e. over 20 years).

Our approach within this and the subsequent sections consists in evaluating key structural features of the real-world Inter-firm Trading Network by normalising, comparing or contrasting to a "randomised network". The latter being synthetically built through swapping links randomly while preserving the degree distributions of each of the firms\cite{Milo:2003}. In a conceptually similar manner to other studies, we opted to maintain the degree distributions to keep the basic quantities held by the agents consistent with the network structure\cite{Viegas:2020:Allometric}. This approach is required since different quantities within agents (i.e. annual sales, number of employees, number of business connections) scale in a similar manner to the power law like degree distribution of companies. Therefore, consistency among these attributes can only be maintained if degree distributions are kept fixed.

\begin{figure}[b!]
  \begin{center}
    \includegraphics[width=1.1\hsize]{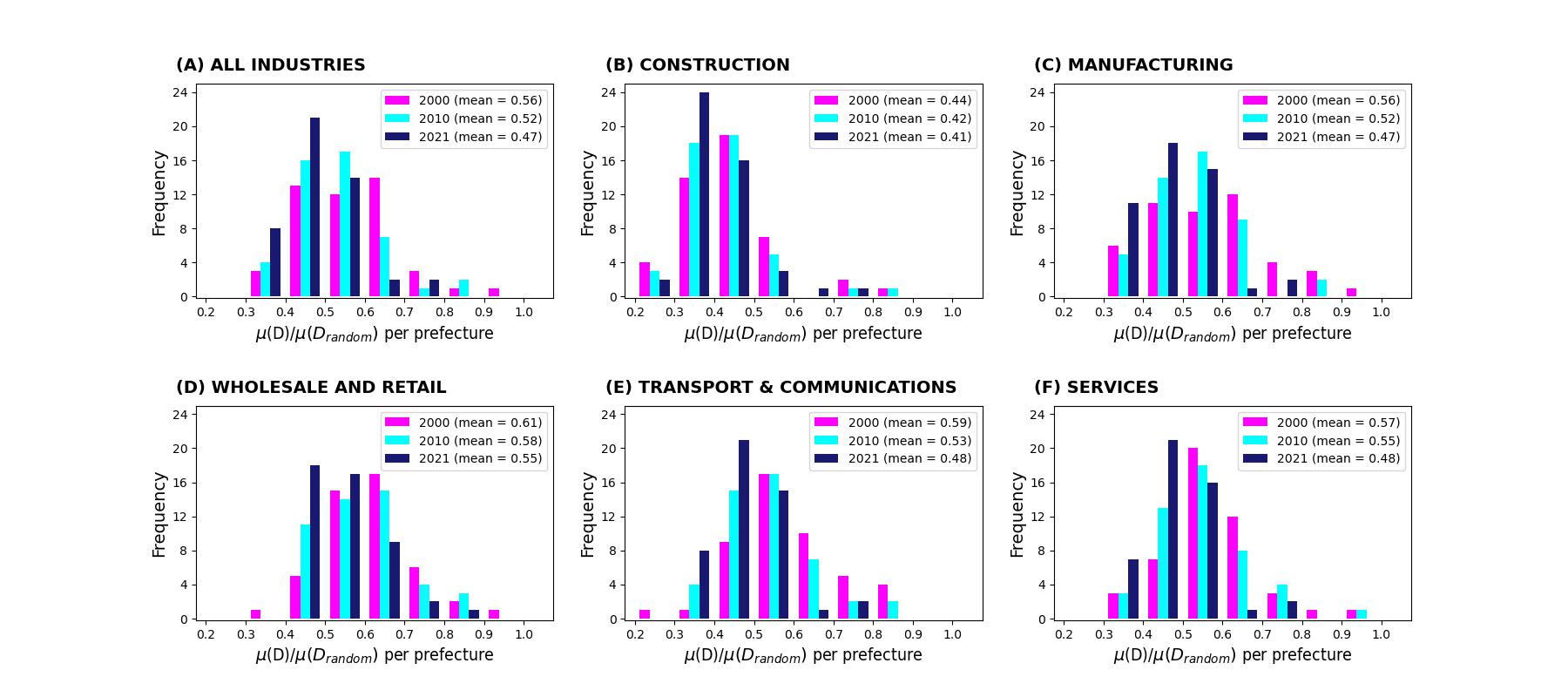}
    \caption{ {\bf Trading proximity is decreasing with the years in all sectors.} \
      Industry sector analysis for geometric proximity between interacting firms, clustered by prefecture for 2000, 2010 and 2021.\
      Each plot shows a histogram, where each prefecture is an item, for different bands $\mu(D)/\mu(D_{random})$ as calculated by the mean of trade distance within the real network $\mu(D)$ and the equivalent values in the randomised network $\mu(D_{random})$.\
      This is done for ({\bf A}) all industries, and the following economic sectors: ({\bf B}) construction industry, ({\bf C}) manufacturing industry, ({\bf D}) wholesales and retailing industries, ({\bf E}) transport and communications industries, and ({\bf F}) services industry. \
      Magenta, cyan and navy blue bars in each plot related to the years 2000, 2020 and 2021, respectively. \
      Moreover, the overall mean ratio for each year is shown within the insets for each plot. 
      \label{fig:2}}
  \end{center}
\end{figure}

Figure \ref{fig:2} shows how geometric proximity between interacting firms behaves by industry sector and coarse grained by prefectures for the years 2000, 2010 and 2021. Each subplot is a histogram of the frequency distribution of the Japanese prefectures as a function of the ratio $\mu(D)/\mu(D_{random})$ for years 2000 (pink), 2010 (cyan) and 2021 (navy blue) with the overall mean ratio for each year shown within the insets. The detailed methodology for the analysis is described in subsection~\ref{meth:industry}.
These results suggest that the average trade distance for the real network $D$ is - and has historically been - consistently and considerably shorter than those controlled randomised network $D_{random}$. 

Such observation is obviously consistent with our previous findings within Fig.~\ref{fig:1}, and it methodologically corroborates the empirical observation that firms tend to be influenced by their business partners distances when forging trade links. In addition, this analysis adds to further pieces of factual knowledge. 

Firstly, although some differences can be observed when comparing different industry sectors, these tend to be limited in nature. The construction sector (panel ({\bf B})) - with the smallest ratio $\mu(D)/\mu(D_{random})$ - tends to be largest outlier, whereas all the others, including the largest (wholesale and retail sector) tend to be fairly similar to each other.

Secondly, and interestingly, the trend over the last 21 years is that of shortening of trade distances within the real world network when normalised to the randomised network. This is a feature that can be observed in all sectors, and it is highly consistent with the theory of the growth and scaling of cities and urban environments\cite{West:2007}. 

Furthermore, we tested whether such a decrease in trading proximity with the years is related to previously existing companies or to newcomers (i.e. the growing number of new companies in each sector). This was done through reproducing the same method but only keeping links related to companies that existed in all years (i.e. 2000, 2010 and 2021). We found that the shortening of distance is related to both existing and new companies in all industries, with the sole exception of the Transport and Communications sector where the reduction has been solely driven by newcomers.

We have also analysed whether potentially different patterns may exist as a function of the size of firms. This was done by (a) splitting the data from the analysis in Fig.~\ref{fig:2} into two subsets related to the size of firms, where the annual sale for both customer and supplier firms are either $S < 10^5$ or  $S \geq 10^5$; and (b) computing the mean trade distances within each subset of all edges for a given prefecture and industry for the years 2021 and 2000 (noting that all prefectures are given the same weight). We found that the average rate of decrease in trading distances over 21 years between the largest companies ($S \geq 10^5$) range $10\%$ to $24\%$ (by industry sector) which is generally the double of that between smaller companies ($S < 10^5$) that vary $5\%$ to $10\%$. This is a trend observed for all sectors, except for Transport and Communications where behaviours are similar regardless of size, revealing again to be contrarian sector.

\subsection{Location Dependency of Trade Distance Distribution: Prefectures and Economic Regions}
\label{sub:ld}

Statistical characteristics of trade distance within a country will be naturally affected by the geographical shape of a given country. For example, trade distance within Japan - an archipelago resembling a bow-shaped form - as opposed to France - a rectangular shaped country - will fundamentally differ simply as a function of geographical boundaries.
Therefore, normalisation procedures are required in order to reduce and minimise the above effect. For the analysis of trade distance distributions, such normalisation can be done by making use of a randomised network.

\begin{figure}[b!]
  \begin{center}
    \includegraphics[width=1.0\hsize]{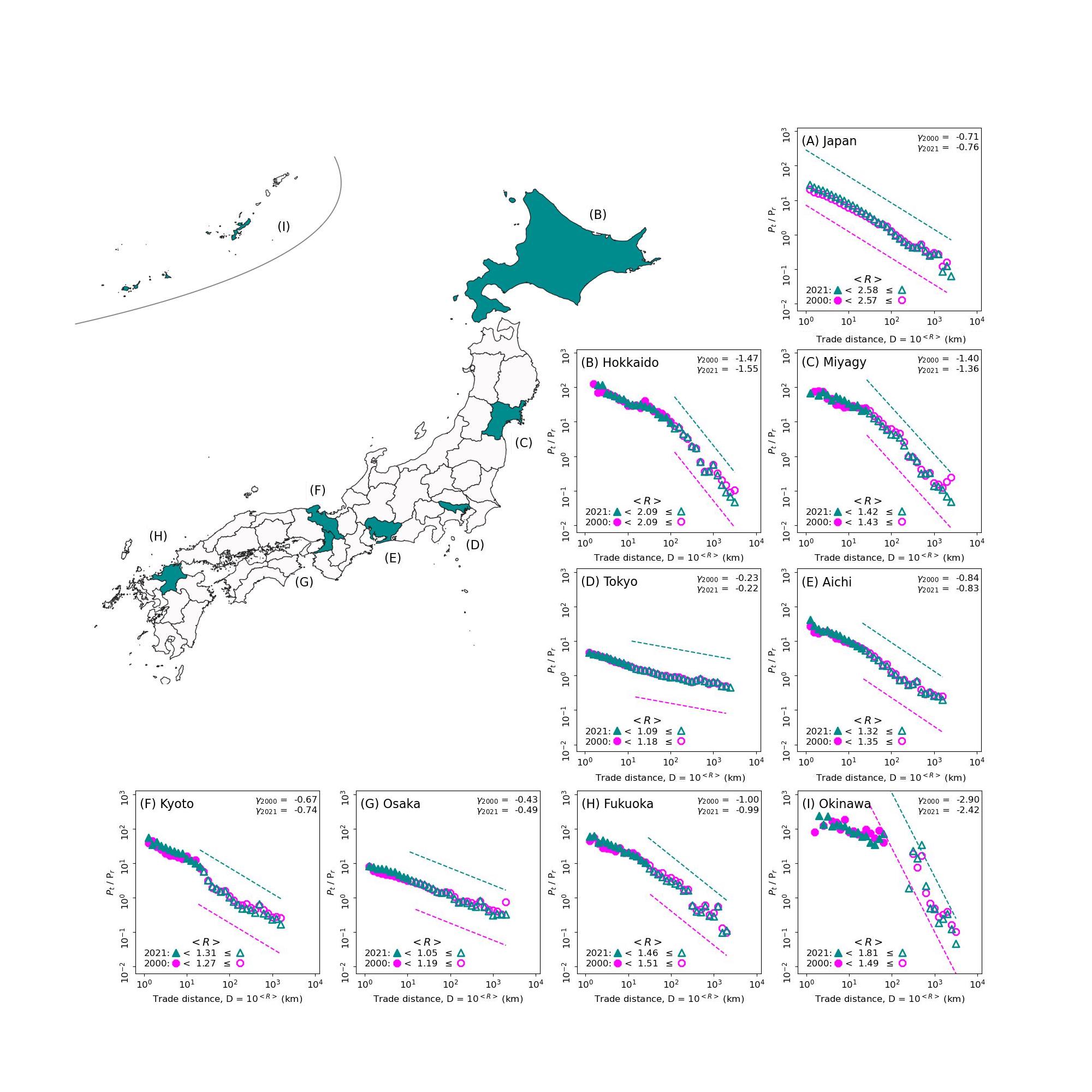}
    \caption{ {\bf Trade distances generally follow a truncated power law at prefecture level.} (Generated by QGIS 3.6 software\cite{QGIS}). \
      Normalised probability distributions of trade distance for each selected prefecture in 2000 and 2021, on a log-log scale plot\
      Each curve of filled magenta square, empty magenta square, filled dark cyan circle and empty dark cyan circle shows the distribution below and above the radius for each prefecture in 2021 and 2000, respectively. \
      The dotted lines indicate power-law distributions. \
      Each panel corresponds to the distribution generated by firms located in ({\bf A}) Japan, and ({\bf B}) Hokkaido, ({\bf C}) Miyagi, ({\bf D}) Tokyo, ({\bf E}) Aichi, ({\bf F}) Osaka, ({\bf G}) Kyoto, ({\bf H}) Fukuoka, and ({\bf I}) Okinawa prefectures, respectively.
      \label{fig:3}}
  \end{center}
\end{figure}

Here, we normalise the probability distribution $P_{t}$ of the of the link distances of a firm in the real-world Inter-firm Trade Network by divideing each number by its equivalent probability distribution $P_{r}$ for the randomised network. In this way, we obtain the resulting normalised probability distribution $P_{t}/P_{r}$ within the panels ({\bf A}) to ({\bf I}) in Fig. \ref{fig:3}. The detailed methodology is described in subsection~\ref{meth:prefec}. For each of these panels, the existence of a well approximated power-law decay above a certain distance threshold can be observed, with the exponent $\gamma$ being estimated by a process similar to the Castillo-Puig test\cite{Castillo:1999,Pisarenko:2011}. In order to detect a sensible starting point of the power-law decay quantitatively, we calculate each size of a given economic zone by the firms' location $(x_{i},~y_{i})$, in accordance with Eqs.~\ref{eq:econon} and \ref{eq:1}. Each $\overline{R}$ and $\left<R\right>$ represents the centre of gravity and the radius of the selected prefecture, respectively.

In Figure \ref{fig:3} we can see the normalised probability distributions of the trade distance in ({\bf A}) Japan and by firms solely located in distinct prefectures: ({\bf B}) Hokkaido, ({\bf C}) Miyagi, ({\bf D}) Tokyo, ({\bf E}) Aichi, ({\bf F}) Osaka, ({\bf G}) Kyoto, ({\bf H}) Fukuoka, and ({\bf I}) Okinawa.
Filled shapes (circles and triangles) show datapoints at a distance $D < 10^{\left<R\right>}$ (as defined in Eq. \ref{eq:1}), whereas empty shapes, present datapoints above the radius, where the power-law decay can be effectively observed.
We note that magenta circles, representing data for 2021, almost always sit on the top of the dark cyan triangles, related to 2000. Therefore, it is reasonable to conclude that little changes have occurred to the probability distributions of first at a high, macro, level. 
Moreover, the power-law decay above a certain distance is consistent with the statistical findings of previous studies on infrastructure networks, such as subway and logistics\cite{Rozenfeld:2002,Yook:2002,Brunet:2002,Marc:2003,Barrat:2004,Guimera:2005,Jung:2008,Marc:2011}.
We also note that the decay exponents significantly differ between prefectures.

\begin{figure}[t!]
  \begin{center}
    \includegraphics[width=1.0\hsize]{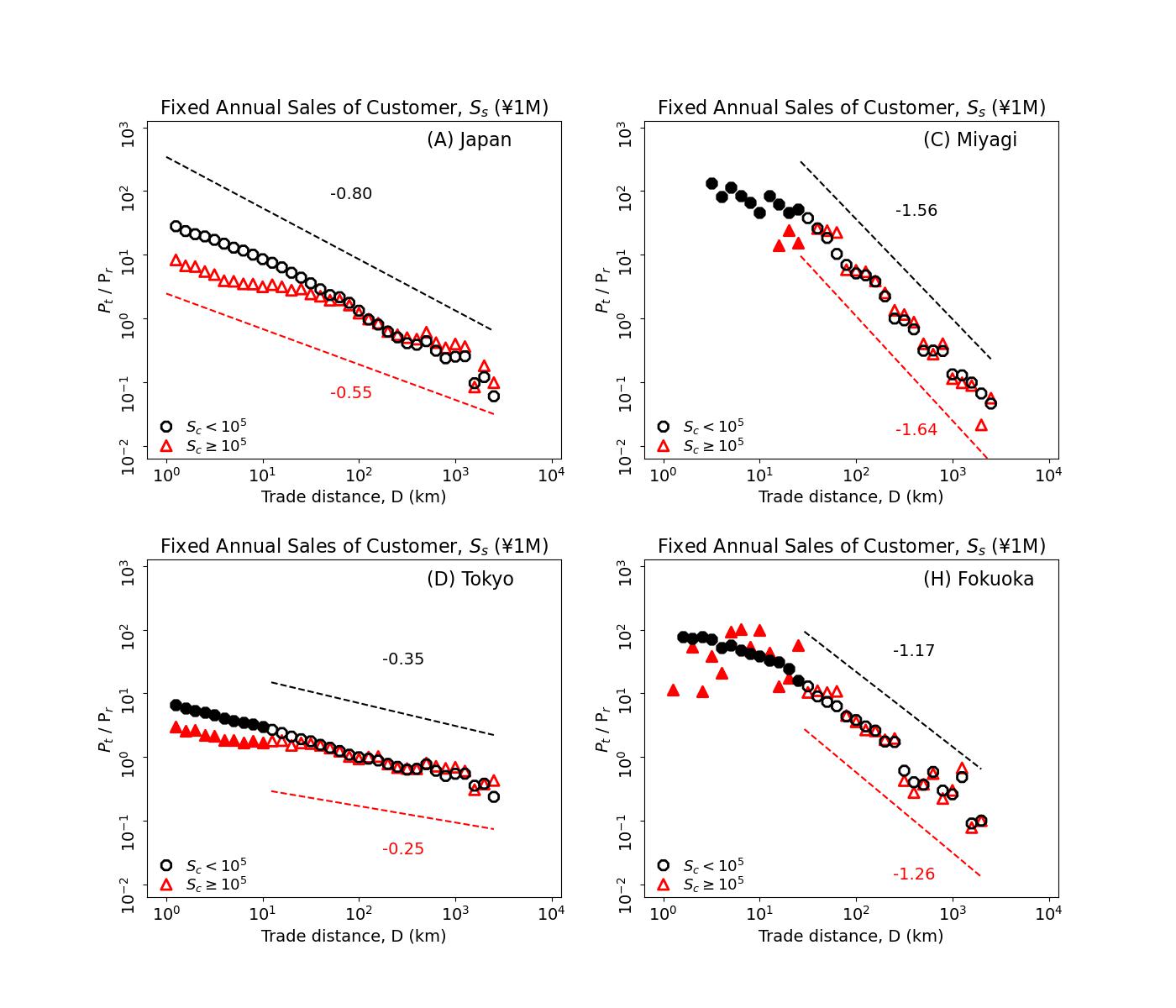}
    \caption{ {\bf Firms of different size closely resemble each other at longer distances, but have very distinctive patterns at shorter distances.} \
       Normalised probability distributions of trade distance for selected prefectures at fixed annual sales of customers (million yen) in 2021, on a log-log scale plot.\
       Each curve of filled shape shows the distribution below the radius $D = 10^{\left<R\right>}$ km for each prefecture in 2021, whereas non-filled shapes are datapoints above the radius. \
      The dotted lines indicate power-law distributions. \
	  Each panel contains the normalised probability distributions $P_{t}/P_{r}$ for ({\bf A}) of Japan as a whole; ({\bf D}) Tokyo, ({\bf C}) Miyagi, and ({\bf H}) Fukuoka.\
      Each curve of black circles and red triangles shows the normalised probability of existence of a trade within distance $D$ in 2021, categorised by the amount of annual sales of companies as $S < 10^5$ and $S \geq 10^5$, respectively. \
     \label{fig:4}}
  \end{center}
\end{figure}

Furthermore, it is also important to note that the shapes for urban areas, such as Tokyo and Osaka in panels ({\bf D}) and ({\bf F}) of fig \ref{fig:3}, are much flatter than others.
This effectively means that location of firms within urban area tends to bear little relation to transaction costs, if one were to assume the direct link between such costs and trade distances as advocated by Coase's theory of firms\cite{Coase:1937}. However, the Coase's theory may be too simplistic to the modern age of human conurbations. 
In order to get more fine-grained insights into the specific dynamics of firm-scale dependencies between and across small, middle-sized and large firms, we generated Fig.~\ref{fig:4}. The plots within the figure are similar in structure to those of  Fig.~\ref{fig:3}. They are distinct, however, by spplitng business transactions (or edges) by the threshold levels (from Fig.~\ref{fig:1}) of the customer annual sales, $S_{c} < 10^5$ (black circles) and $S_{C} \geq 10^5$ (red triangles). 
Each panel contains the normalised probability distributions $P_{t}/P_{r}$ for ({\bf A}) of Japan as a whole; ({\bf D}) Tokyo, the largest prefecture; ({\bf C}) Miyagi, a small prefecture; and ({\bf H}) Fukuoka, a middle sized prefecture. 

The four panels clearly indicate that the normalised probability distributions $P_{t}/P_{r}$ of firms both above (red triangles, $S_{C} \geq 10^5$) as well as below (black circles, $S_{c} < 10^5$) threshold closely resemble each other for all prefectures at longer distances, but they have very distinctive patterns at shorter distances.
By computing the best fitting for the slope (in the same manner as done for Fig.~\ref{fig:3}), it is possible, however, to observe that, quantitatively, the similarity increases from the largest to the smallest prefectures. Here, however, it is important to emphasise that the starting point of the fitting curve (radius $D = 10^{\left<R\right>}$ km) has some impact on the exact value of the slope. The described tendency, however, is still valid, but less pronounced, if the starting point is moved to higher distances. 
Most important though is the fact that larger companies tend to be much less sensitive to distance at shorter levels than smaller companies. This behaviour can be noticed in all panels within Fig.~\ref{fig:4} as the red triangles are always below the black circles at shorted distances (left side) and have a much flatter pattern. 
Assuming some level of validity for Coarse's theory of firms, and by combining our observations in fig \ref{fig:3} and \ref{fig:4}, we are able to conclude that, at shorter distances, small and middle-scale firms might have strong concerns about transaction costs than large scale companies. However, the gap between these behaviours tend to disappear at longer distances.

It is reasonable to hypothesise that these observed dynamics may well be influenced by the high-quality infrastructures in the urban areas, since most large companies are located in the core economic centres of Tokyo and Osaka.
Moreover, it is important to note that the major benefits from improvements in accessibility such as High-Speed Rail\cite{Brocker:2010,Monzon:2013} may have had a "democratic" effect for firms across the size spectrum. In this regard, the Japanese Shinkansen, which started activities in 1964, may be seen as a truly beneficial transport policy equally benefiting all firms.

\subsection{Prefectures and Communities: A Mutual Information Approach}

\label{find:mutual}

The previous section highlighted distinctive features among prefectures having different urban density, economies and sizes. The boundaries of prefectures, however, are set upon historical administrative and geopolitical structures that might not always resemble the real trading partnerships arising from close associations in economic structures. 
Within this in mind, we developed a method to aggregate, or coarse-grain, prefectures into economic regions based on the existing real world interactions within the Inter-firm Trade Network. We then follow by comparing and contrasting the resulting distribution of the geometrical trade distances for each of the regions.

Here we summarise the logic and rationale for the application of our method from a theoretical framework and empirical perspective. The detailed method and specific steps in developing our analysis are described within subsection~\ref{meth:mut}.

It is now well known that the distribution of nodes and edges for the Japanese Inter-firm Trade Network follows a power law distribution governed by mechanisms associated with a cumulative advantage [16] and preferential attachment [17], leading to the formation of a disassortative network underpinned by a power law structure [18].
Previous studies show that, from an information theory perspective, an amount of mutual information that can be calculated as a function of source and target pairings will always be different from zero if the network is disassortative. Therefore, within our framework (where customer and suppliers are the equivalent to source and target), it is possible to break the computation of the mutual information into two separate but related components: The structural mutual information, $SMI$, and the total mutual information, $I$. The former solely relates to the degree distribution of the nodes within a given network, whereas the latter encompasses both the node degree distribution as well as the disassortativeness of the network. In our research, $SMI$ is calculated from the randomised network, whereas $I$ is calculated from the real world Inter-firm Trade Network.
Such distinction also fits well within the economics and finance perspective since $SMI$ is closely related to a theoretical ’free-market’, stock market-type configuration, where a buyer has a probability to trade with a seller solely based upon the existing quantities of stock held by the latter. In contrast, $I$ reflects the natural market distortions associated dominance and influence with a real world situation, where "preferences" of buyers and sellers are expressed.

\begin{figure}[b!]
  \begin{center}
    \includegraphics[width=1.1\hsize]{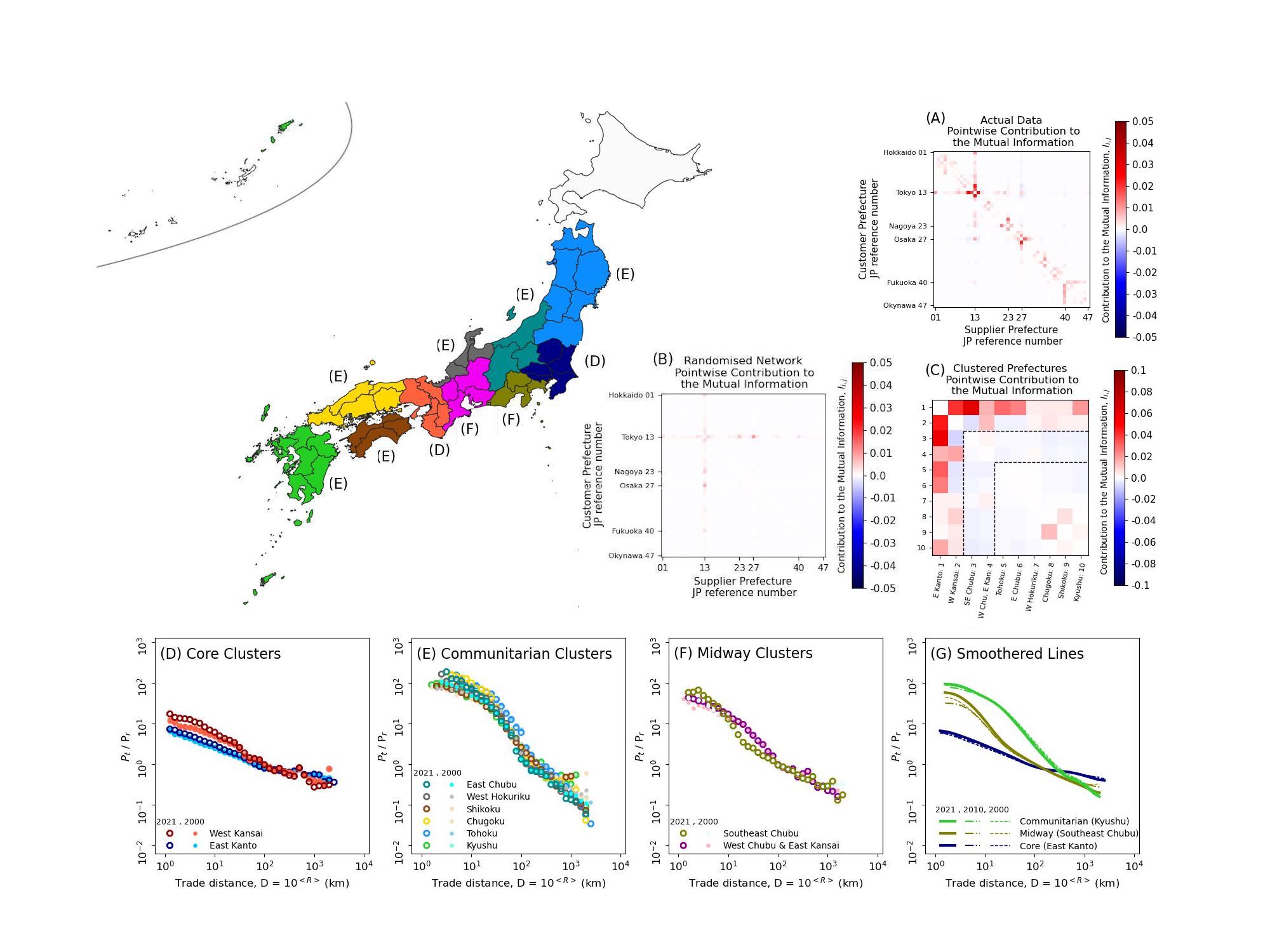}
    \caption{ {\bf Pointwise Contribution to the Mutual Information and Structural Clustering of Prefectures} (Generated by QGIS 3.6 software\cite{QGIS}). \
      The map on the left is coloured in accordance with the eight regions that emerge from the clustering method based on the pointwise positive contribution to the mutual information.\
      The regions are aggregated into three groupings: The Communitarian Clusters ({\bf E}), corresponding to Kyushu (lime green), Chugoku (gold), Shikoku (saddle brown), Tohoku (dodger blue) and East Chubu (dim gray); The Core Clusters ({\bf D}) of Tokyo Sphere (navy blue) and Osaka Sphere (dark orange); and the Midway Cluster ({\bf F}) (dark magenta).\
      The panels at the bottom row show the normalised probability distributions $P_{t}/P_{r}$ with distance (similarly to Fig.~\ref{fig:4}), where Panel ({\bf G}) being a schematic trend-line representation of the three groupings.\
      Panels ({\bf A}), ({\bf B}) and ({\bf C}) present the heatmaps for the pointwise contribution to the mutual information (calculated in accordance with Eq.~\ref{eq:mif}) for the real world Inter-firm Trade Network, the randomised network and the clustered regions, respectively.
      \label{fig:5}}
  \end{center}
\end{figure}

Panels ({\bf A}) and ({\bf B}) of Fig.~\ref{fig:5} show, respectively, the pointwise contribution to the mutual information for the real world Inter-firm Trade Network and randomised Network for the parings of customers (vertical) and suppliers (horizontal) by prefecture, coarse-grained in accordance with Eq.~\ref{eq:coarse}. From analysing the structure of Eq.~\ref{eq:mif}, one can deduct that a positive pointwise contribution indicates that the pairing occurs in a frequency `higher than expectations', whereas a negative contribution occurs with frequency `below expectations'. In the context of a trading relationship, this can be interpreted as the prefectures having a higher level of attraction between the customer and supplier for the former whereas the later shows a tendency for repealing each other.
From ({\bf A}), three key features can be observed. Firstly, the more urbanised the prefecture (Tokyo, Osaka, Nagoya, Fukuoka) the higher is the tendency to record a positive pointwise contribution to the mutual information. Secondly, as the cardinal order of prefectures are closely related to geographical proximity, it can be clearly noticed that neighbouring prefectures have a higher frequency of positive pointwise contribution (as identified visually by the shapes around the diagonal). Thirdly, the relation among prefectures (specially Tokyo) tends to be relatively symmetrical (visually, horizontal lines similar to vertical lines).
In contrast, it is possible to verify from ({\bf B}), that a randomised network would have kept some level of mutual information, the $SMI$, mainly at level of Tokyo only, the largest prefecture, with the diagonal shapes associated with neighbourhood been effectively eliminated.

Next, we apply a clustering method where we aggregate into single trade regions all entities that have positive (in both directions) contribution to the mutual information among themselves. The clustering process start identifying the largest possible number of prefectures that can be fit into a single cluster with the highest levels of aggregated pointwise contribution to the mutual information $I_{ij}$, segregating it, setting it aside, and repeating the process with the remaining prefectures until a cluster can no longer be found. Any remaining unallocated prefectures (in our case, only Wakayama and Tochigi) is then allocated to a cluster if it contains fewer than two negative contributions $I_{ij}$ between the prefecture and the cluster. The resulting clusters can be observed from the map in Fig.~\ref{fig:5}. 

From the analysis, a number of interesting observations can be made.

Firstly, even though no geographical feature is enforced at input level, all clusters are formed within actual geographical neighbours, and only the remote islands of Hokkaido and Okinawa remain unclustered.

Second, four clusters emerge as an exact match of Japan's geographically defined regions, namely: Kyushu (lime green), Chugoku (gold), Shikoku (saddle brown)and Tohoku (dodger blue). These clusters, together with a smaller ones that we named East Chubu (darkcyan) and West Hokuriku (dim gray), make up the grouping we define as `Communitarian Clusters'. 
Importantly, as it can be noted in panel ({\bf E}), all Communitarian Clusters exhibit very similar characteristics in terms of normalised probability distributions $P_{t}/P_{r}$. It is possible to argue, therefore, that from a structural trading perspective, these clusters have remarkable similarity, which cannot be observed at single prefecture, administrative, level.

Third, we note that a second group, the `Core Clusters', emerge from the two urban regions associated with the two largest cities in Japan, Tokyo (East Kanto) and Osaka (West Kansai). These two clusters also show similar patterns in relation to the normalised probability distributions $P_{t}/P_{r}$ (panel ({\bf D})), although some limited divergence exists at shorter trade distance levels.
In the previous subsection, \ref{sub:ld}, we commented on the distinct patterns arising between urban prefectures and those with lower density. By comparing and contrasting, panels  ({\bf D}) and  ({\bf E}), it is possible to get a much enhanced picture where the Core (urban) Clusters have dynamics fundamentally different from the Communitarian Clusters. Whereas the former tends to have higher tendency of trade across the nation (i.e. it is flatter), the latter shows a clear regional, limited geographical, tendency (i.t. the decay is very pronounced).

Fourth, a third grouping, the `Midway Clusters', encompass two smaller clusters, namely South East Chubu and West Chubu \& East Kansai. The prefectures within these clusters are characterised by higher prevalence of negative contributions to the mutual information beyond Tokyo.
It is also worthy to note that the midway clusters are geographically located between the two largest urban centres, and that the normalised probability distributions $P_{t}/P_{r}$ for the clusters have hybrid shape with characteristics close to the Communitarian Clusters at shorter distances and more similar to the Core Clusters at longer distances (as it can be observed in panels ({\bf F}) and  ({\bf G})).

Fifth, panel ({\bf C}) shows the computation of the pointwise contribution to the mutual information coarse grained at cluster levels, where three general and distinct patterns can be clearly noticed depending on the grouping. The Core cluster of Tokyo is generally characterised by a significant positive contribution across most of the clusters, whereas the Osaka cluster have positive contributions with clusters Central to West Japan, but negative nearing Tokyo and North Japan. In contrast, the Midway Clusters tend to be negative in relation to the Communitarian clusters - and between themselves - but positive in relation to East Kanto (Tokyo) cluster. Finally, the Communitarian Clusters indicate very little level of contribution among themselves, albeit with some moderate positive contribution among the neighbouring far West clusters. Putting simply, East Kanto has a nationwide `attraction', West Kansai `attracts' the Western half of the countries, whereas all others have limited regional reach.


\section{Methods}

All data used in this study is collected and provided by Teikoku Databank, Ltd., one of the largest corporate research providers in Japan. It consists of corporate and business transaction data stretching from 1995 to 2021, generally encompassing over 1 million companies and over 4 million transactions per year. Whereas we carried out our analysis for all years within the above range to ensure the consistency of our findings, we have chosen to present our results based on years 2000, 2010 and 2021 as these are good representations of the evolution of the Inter-firm Trade Network over time as well as its core structural features.

\subsection{Scaling Relations between Interacting Firms}
\label{meth:scaling}

In this subsection, all references to panels related to those within Fig.~\ref{fig:1}.

\

An individual business transaction is represented by an edge $e_{ij}$ within the set of all transactions $E$. Two firms, a Customer $i$ and a Supplier $j$ (within the set of firms $F$), form pair of nodes within a given edge. 

The geographical, or trade, distance $D_{ij}$ for each edge equates to the Euclidean distance of the Cartesian latitude and longitude coordinates $(i_{lat},i_{lon})$ and $(j_{lat},j_{lon})$ for the pair $i$ and $j$, respectively. By construct, it follows that $D_{ij} = D_{ji}$, however $e_{ij} \neq e_{ji}$.  Therefore,

\begin{eqnarray}
  \epsilon : \{e_{1,1},e_{1,2},...,e_{i,j}\} \rightarrow \{D_{1,1},D_{1,2},...,D_{i,j}\}  \label{eq:10}
\end{eqnarray}

Within Panels({\bf A}) and ({\bf B}), individual edges $e_{ij}$ are aggregated into groups $B_{m,n}$ based on the annual sales $S$ of the firms $i$ and $j$,

\begin{eqnarray}
  B_{m,n}  = \{e_{ij} \  | \  km < \log_{10}(S_{i}) < k(m+1)\ , \ kn< \log_{10}(S_{j}) < k(n+1) \ , \ e_{ij} \in E \ \} 
   \label{eq:11}
\end{eqnarray}
\noindent where $k = 1, m = \mathbb{N}$ and $n = \mathbb{N}$.

\
 
Panel ({\bf A}) represents the cardinality $|B_{m,n}|$ whereas panel ({\bf B}) relates to the median of the trade distance $\widetilde{\epsilon({B_{m,n}})}$  for each group.
\\
 
The edge aggregation for panel ({\bf C}), where $S=S_{i}$, and panel ({\bf D}), where $S=S_{j}$, follows the construct,

\begin{eqnarray}
  B_{m,c}  = \{e_{ij} \  | \  km < \log_{10}(S) < k(m+1) \ , \ c \ , \ e_{ij} \in E \ \} 
   \label{eq:12}
\end{eqnarray}

\noindent where $k = 0.3$, $m = \mathbb{N}$, $c = \{S < 10^{5}$ (black circles) , $S \geq 10^{5}$ (red triangles)\}.

\subsection{Geometric Proximity between Interacting Firms: Industry Sectors and Prefectures}
\label{meth:industry}

Single edges $e_{ij}$ within the set $E$ are aggregated by paring groups $B_{pref,ind}$ of prefectures and industry categories.

\begin{eqnarray}
  B_{pref,ind}  = \{e_{ij} \  | \  pref,ind \in \{(Pref_{i},Ind_{i}),(Pref_{j},Ind_{j})\} \ , \ e_{ij} \in E \ \} 
   \label{eq:21}
\end{eqnarray}

In a similar manner to the edges $E$ for real world network, the trade distances $D_{ij}^{rand}$ are calculated for each of the edges $e_{ij}^{rand}$ from the randomised network $E^{rand}$ with the corresponding functional mapping $\epsilon^{rand}$, Eq.~\ref{eq:10}. The randomised network $E^{rand}$ is also clustered in accordance with Eq.~\ref{eq:21}, resulting in equivalent groups $B_{pref,ind}^{rand}$.

For each combination $(pref,ind)$, the ratio of the mean distances $\mu(D)/\mu(D_{random})$ between the real world and randomised network is then calculated

\begin{eqnarray}
  \frac{\mu(D)}{\mu(D_{random})} \ _{pref,ind}  = \frac{\mu(\epsilon(B_{pref,ind}))}{\mu(\epsilon^{rand}(B_{pref,ind}^{rand}))} \ 
   \label{eq:22}
\end{eqnarray}

Each panel of Fig.\ref{fig:2}, therefore, fixes the industry category and constitutes a histogram of the ratio for each prefecture, as calculated by Eq.~\ref{eq:22}. 


\begin{eqnarray}
B_{pref,ind,\geq 10^{5}}  = \{e_{ij} \ |  \ S_{i} \geq 10^{5} \ , \ S_{j} \geq 10^{5} \ , \ e_{ij} \in B_{pref,ind} \ \} \\ B_{pref,ind,< 10^{5}}  = \{e_{ij} \ |  \ S_{i} < 10^{5} \ , \ S_{j} < 10^{5} \ , \ e_{ij} \in B_{pref,ind} \ \}
   \label{eq:23}
\end{eqnarray}
 
It follows that each datapoint within each panel represents the Cartesian coordinates $\mu(\epsilon(B_{pref,ind,f_{size}}))$ for the years 2021 and 2000 respectively.

\subsection{Location Dependency of Trade Distance Distribution: Prefectures and Economic Regions}
\label{meth:prefec}

Individual edges $e_{ij}$ with the set $E$ are aggregated into groups $B_{pref,d}$ based on prefectures and related trade distances $D_{ij}$,

\begin{eqnarray}
  B_{pref,d}  = \{e_{ij} \  | \  pref \in \{Pref_{i},Pref_{j}\} \ , \  kd < \log_{10}(\epsilon(e_{ij})) < k(d+1) \ , \ e_{ij} \in E \ \} 
   \label{eq:41}
\end{eqnarray}
\noindent where $k = 0.1, m = \mathbb{N}$.

The cardinality $|B_{pref,d}|$ for each set is therefore the frequency of point $d$ with the distribution frequency $\phi_{pref}$ which is in turn turned into a probability density $P_{pref}$. 

The same method within Eq.~\ref{eq:41} and thereafter is then applied to the randomised network set $E^{rand}$, ultimately resulting a probability density $P_{pref}^{rand}$. The ratio $P_{pref}$/$P_{pref}^{rand}$ form therefore the data for Fig.~\ref{fig:3} (noting that any probability density under 0.01\% is excluded). 

Fig.~\ref{fig:4} follows a similar process for the selected prefectures with the distinction that all subsets $B_{pref,d}$ and $B_{pref,d}^{rand}$ are split into two subsets, conditional to $S_{i} < 10^{5}$ or $S_{i} \geq 10^{5}$.

Each size of a given economic zone by the firms' location $(x_{i},~y_{i})$, expressed in longitude and latitude of form $i$, is calculated as follows:
\begin{eqnarray}
\label{eq:econon}
  \overline{R} &=& \left(\overline{x}, \overline{y}\right)= \left(\frac{1}{N} \sum_{i=1}^{N} x_{i}, \frac{1}{N} \sum_{i=1}^{N} y_{i}\right), \\
  \left<R\right> &=& log_{10}\left(\frac{\tau}{N} \sum_{i=1}^{N} \sqrt { \left(x_{i}-\overline{x}\right)^{2} + \left(y_{i}-\overline{y}\right)^{2} } \right),
  \label{eq:1}
  \label{eq:2}
\end{eqnarray}
where $N$ is the total number of firms and $\tau$ is the conversion factor from the geographical degree to the metrical distance (km).

\subsection{Prefectures and Communities: A Mutual Information Approach}
\label{meth:mut}

Individual edges $e_{ij}$ within set $E$, are coarse-grained by prefectures as (customer,supplier) pairings as follows

\begin{eqnarray}
    L_{Pref_{i},Pref_{j}} = \{e_{i,j} \  | \  (i_{pref},j_{pref}) = (Pref_{i},Pref_{j}) \ , \ e_{ij} \in E )\} \}
    \label{eq:coarse}
\end{eqnarray}

The probability distributions $P_{ij}$ for each pairing of prefectures, the probability distribution of a prefecture being a Customer $P_{i}$ and the probability distribution of a prefecture being a Supplier $P_{j}$ are:

\begin{eqnarray}
    P_{ij} = \frac{|L_{ij}|}{|E|} \  ,  \  P_{i} = \frac{\sum_{x=1}^{47}{|L_{ix}|}}{|E|} \  ,  \  P_{j} = \frac{\sum_{x=1}^{47}{|L_{xj}|}}{|E|}
\end{eqnarray}

\noindent noting here that the Japanese prefecture are given specific numbers, and therefore, $i$ and $j$ are a range [1..47].

The pointwise contribution to the mutual information for each pair of prefectures is therefore 

\begin{eqnarray}
    I_{ij} = P_{ij}\log2(\frac{P_{ij}}{P_{i}P_{j}}) 
    \label{eq:mif}
\end{eqnarray}


\section{Discussion}

As detailed within the \textit{Introduction} section, we designed this research to identify the core emerging properties of the inter-firm business transaction networks by making use of a large, extensive and comprehensive dataset that materially encompass all transactions within Japan.

Unsurprisingly, we found that there is a generic functional relationship between geometric distance and frequency of trades which is largely dependent on the size of the firms within a customer and supplier relationship. 
Surprisingly, however, is the fact that whereas such relationship scales uniformly and symmetrically among smaller and middle size companies, it breaks down when large companies are involved. Essentially, small firms will go large distances to partner with big suppliers. Yet large firms will not go long distances to partner one another, as illustrated by Fig. \ref{fig:1}. 

A second finding that is somehow counterintuitive, or against the general expectations, is the fact that trade distances are - on average - reducing over time when, viewed for a relatively long period of $21$ years (between $2000$ and $2021$). There is some level of general wisdom among policymakers that with all the advances of technology and transport during the $21^{st}$ century, economic and business relationships would reduce dependence on geographical proximity. Yet, our results suggest that the trading proximity within the physical world still holds an important role in the optimisation of costs and logistics of companies (as illustrated by Fig. \ref{fig:2}), irrespective of the industry sector. If anything, our results give some evidence that the technological innovations may not have substantially affected the dynamics of scaling and the pace of life in cities\cite{West:2007} observed by researchers at the turn of the century. 

Third, we find that normalised probability distribution of trade distance generally follows a shape that resembles truncated power law distributions both at prefecture as well as trade region levels - as illustrated by Figs. \ref{fig:3}, \ref{fig:4} and \ref{fig:5}. From this narrow perspective, our results are consistent with the statistical findings of previous studies on infrastructure networks, such as subway and logistics, suggesting weaker concerns about transaction costs in shorter distances, which are predominantly in terms of frequency within the urban areas. 
This is seemingly influenced by high-quality infrastructures in the urban area, which leads to an improved efficiency and attractiveness of cities as newly-connected locations with reduced communication, transaction, and costs.

Our work also unveils a subtle facet with regards to the influence of regional characteristics and patterns on the dynamics of trade distance.
Specifically within Japan, we are able to clearly identify three groupings of structural trading dynamics based on their similarity (as illustrated by Fig. \ref{fig:5}): the Communitarian, Core and Midway clusters. 

The first grouping, the Communitarian Clusters, consists of six distinct trade regions that emerge naturally from the our mutual information clustering method, and that are very closely associated with the formally defined Japanese regions. The grouping is characterised by rural, lower density areas where trading activity has a very regional, local bias (with all clusters showing very similar, relatively high, power law decay). 
In contrast, the Core clusters are composed of the two regions around the two largest urban areas in Japan, Tokyo and Osaka where trading distances play a much reduced role (with a much smaller power law decay). The mutual information analysis also shows clearly the distinction between these two clusters: Whereas the cluster including Tokyo has a nationwide influence, Osaka's is mostly concentrated at the Western half Japan.
Lastly, the Midway clusters, located between these two urban core areas, are defined by mixed characteristics, in-between, the Communitarian and Core Clusters. 

Here, we would ask the reader to allow us an element of constrained `licentia poetica' to discuss some of the potential consequences of the observed reduction in trade distances and to speculate on existing options to policy makers. At first sight, a `laissez-faire' market view would conclude that the shortening of trade distances can be seen in a positive light if one were to assume that this is the result of efficient logistics leading to some natural level of optimisation that reduces transport and commuting effort, and by extension, emissions and pollution. However, our view is that such argument is simply naive as these improvements might be simply driven by higher levels of urbanisation instead of optimised cross-country logistics, where the second order effects may well negate, or even worsen, any of the perceived benefits\cite{Perera:2023}. Indeed, some academics argue that middle-sized cities are in existential crisis\cite{Wilkinson:2019} with the emergence of the super-cities\cite{West:2007}. Therefore, we would argue that our research is simply a small step towards a better understanding of the evolutionary dynamics of trading distance, and their different effect on regions. However significant further work is required in the field to ascertain whether optimal methods can be designed to support the challenges associated with climate change (such as emissions) as well as social changes (i.e. higher deprivation, lack of opportunities, ageing population, etc. in smaller and middle sized cities) which should take as much, if not higher, importance as economic costing efficiency.

Whereas we carried out our work specifically on Japan, it would be not unreasonable to suggest that there is a likelihood that similar fundamental dynamics are present in most of other larger countries. However, this remains an object of future study if similar data were to be available.
  
We also note that this study is however constrained by the fact that companies are not broken into smaller subunits such as branches. Therefore, the geographical location of a company is based on the ultimate decision-making location - normally the head office - where the management and governance of commercial transactions tend to take place  (as opposed to the actual point of sale). We emphasise, though, that we do not regard approach as limitation, but imply as a method to draw focus on the decision making core rather than on the process driven activities.

\section{References and Notes}

\section*{Acknowledgments: }

\noindent {\bf General: } 
The authors appreciate Teikoku Databank, Ltd., Center for TDB Advanced Data Analysis and Modeling in Tokyo Institute of Technology for providing both the data and financial support. 

\noindent {\bf Funding: } 
This study is partially supported by 
JSPS KAKENHI (Grant No.18H01656) and 
JST, Strategic International Collaborative Research Program (SICORP) on the topic of ”ICT for a Resilient Society” by Japan and Israel.

\noindent {\bf Author contributions: }
EV performed and designed experiments, interpreted the results, and co-wrote the manuscript. 
OL interpreted the results and co-wrote the manuscript.
SH designed experiments, supervised experimental designs, interpreted results, and reveiwed the manuscript. 
HT designed experiments, supervised experimental designs, interpreted results, and reviewed the manuscript. 
MT conceived the project, designed experiments, supervised experimental designs, interpreted results, and reveiwed the manuscript.

\noindent {\bf Other contributions: }
The authors also express their gratitude to Dr. Hayato Goto who worked on some of the earlier experiments and drafts. 

\noindent {\bf Competing interests: }
The authors declare that they have no competing interests.

\noindent {\bf Data and materials availability: }
The data underlying this study was provided from Teikoku Databank, Ltd. 
Interested and qualified researchers can request access to the data set in the same manner as the authors at the following URL: \url{https://www.tdb.co.jp/service/mail_e/form.jsp}

\end{document}


\maketitle

\section*{The files inlcudes}

Section S1. Data Source Context and Description\\
Section S2. Scaling Relation between Interacting Firms\\
Section S3. Correlation between the exponents of sizes of Economic Zones, ⟨R⟩, and the decay $\gamma$ in the normalised probability distributions $P_{t}/P_{r}$\\
Section S4. Distribution of companies across prefectures in Japan\\

\noindent
Fig f1. The frequency of trade and distances $D$ on a randomised network as a function of the annual sales of customers $S_{c}$ and annual sales of suppliers $S_{s}$.\\
Fig f2. The frequency of trade and distances $D$ highly depend on the annual sales of customers $S_{c}$ and annual sales of suppliers $S_{s}$ for the year 2010\\
Fig f3. Correlation between the size of economic zones $⟨R⟩$ and power law exponent $\gamma$.\\
Fig f4. Distribution of companies by prefecture in Japan 2021.\\
Fig f5. Distribution of companies by prefecture in Japan 2021, split by the annual sales, $S$.\\

\noindent
Table S1. Exponent for the sizes of economic zones ⟨R⟩ and $\gamma$ decay of normalised trade distance distributions for each prefecture in 2021.\\

\pagebreak

\section*{S1: Data Source Context and Description}

Business practices in Japan have some unique features. When building relationships with client or managing counterparty credit risk, Japanese banks and firms tend to gather a significant level of corporate information, commonly making use of professional third-party organisations to collect data and evaluate their potential and existing partners’ credit status. Teikoku Databank Ltd (TDB) is one of the largest corporate research providers in Japan. The company was founded more than 1205 years ago and it has been assessing the credit status of firms since then. TDB's credit research reports is very extensive, including detailed information on the financial statements of firms, their history, business partners, management structure, and banking transactions.

Within this research, we make use of the Japanese inter-firm business transaction networks of money transactions between customers and suppliers developed by TDB and derived from the underlying credit research activities. In addition to transactions, TDB's database also contains the geographical coordinates of virtually almost all firms within their domain. Therefore calculation of the Euclidian geometrical distances among all firms becomes a relatively straight forward computational process.  The database contains the register of about 1.5 million corporations with total annual sales beyond 1.4 trillion yen in 2020. Compared with the reports by the Minister of Finance in Japan and the Statistics Bureau of Japan, the database covers around 84\% of the total number of corporations, and about 98\% of their total annual sales, with over 1 million companies with consistent and reliable information on business transactions and geographical location. 

To our knowledge, no other developed country has such an extensive and granular database held outside Governmental Agencies.


\section*{S2: Scaling Relations between Interacting Firms}

\subsection*{S2A: The Randomised Network}

\begin{figure}[b!]
  \begin{center}
    \includegraphics[width=0.9\hsize]{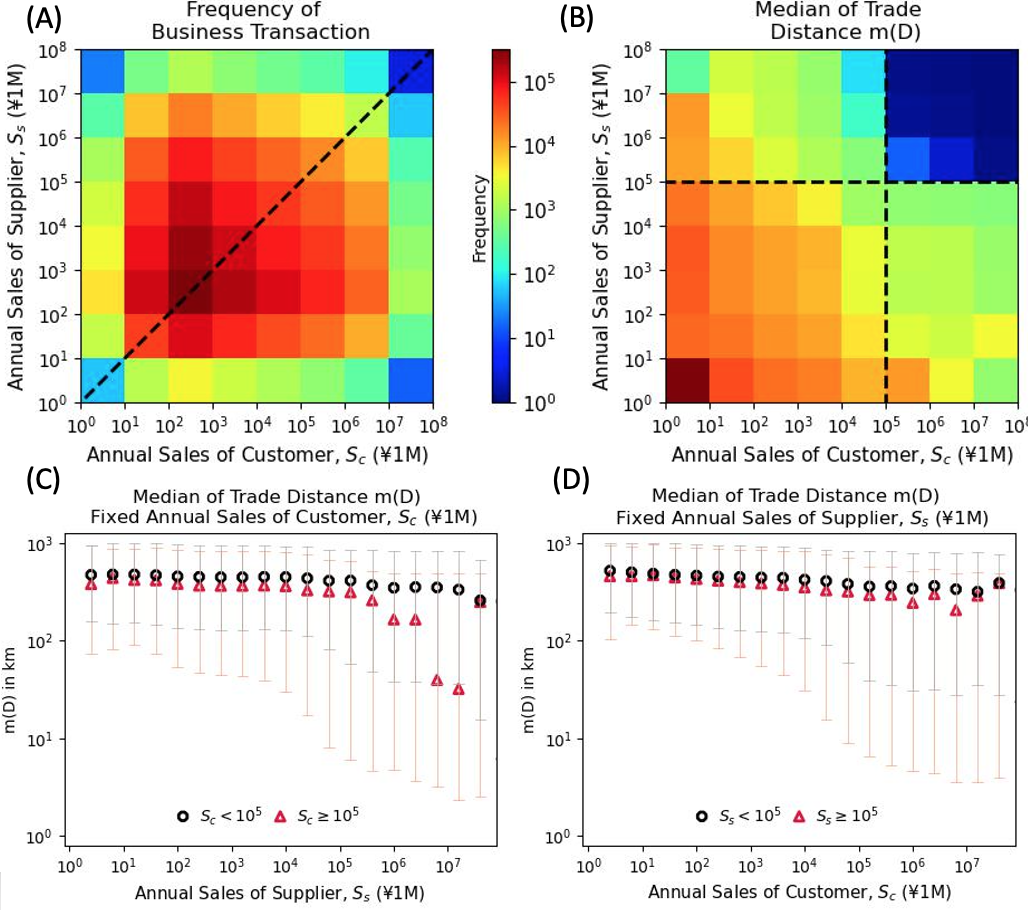} 
    
    \caption{  {\bf The frequency of trade and distances $D$ on a randomised network as a function of the annual sales of customers $S_{c}$ and annual sales of suppliers $S_{s}$.} \
      ({\bf A}) is a heatmap showing the frequency trades among the $S_{c}$ and $S_{s}$ pairings, where the dotted diagonal line is the symmetry $S_{c}$ = $S_{s}$.
      ({\bf B}) is the equivalent heatmap quantifying the the median $m(D)$ of the $S_{c}$ and $S_{s}$ pairing, where the horizontal and vertical lines indicate the original break in symmetry for the actual data, $S_{c} \simeq S_{s}$, which is not found in the randomised network.
      ({\bf C}) shows the scaling of the median distance $m(D)$ of firms at different $S_{s}$ levels (above and below the actual data symmetry break) with fixed $S_{c}$, and\
      ({\bf D}) is an equivalent plot with fixed $S_{s}$ and variable $S_{c}$ levels. \
      Each black circle and red triangle represents firms where $S < 10^{5}$ and $S \geq 10^{5}$, respectively.\
      Figures are plotted on a log-log scale for the year 2021.\
   \label{fig:1s}  
         }
  \end{center}   
\end{figure}

In our (main) research paper paper we show four panels in Fig. 1 to illustrate the scaling relations between customers and suppliers within the Inter-firm Trade Network for the actual data in 2021. Here, we reproduce in Fig. \ref{fig:1s} the same graphs for the data related to randomised network, which was generated by the process described in the Methods section. 

As it can be noted, the randomisation process generates results consistent with expectations.

First, and in contrast to  the actual data, the randomised data is virtually symmetric for both the frequency of trades as well as the median trade distance.

Second, the median trade distance significantly increase on average as the geographic scaling removed. This can also be observed by the panels at the lower row.

Thirdly, it can be clearly noticed that even after normalisation, the  largest firms retain a distinct pattern from the smaller firms. This is simply as a result of the fact that the network is dissortative, and the number of edges are discrete - natural - numbers. In our research paper we explain this feature when elaborating on the concept of the Structural Mutual Information, SMI.

\subsection*{S2B: The Actual Data for 2010}

\begin{figure}[b!]
  \begin{center}
    \includegraphics[width=1\hsize]{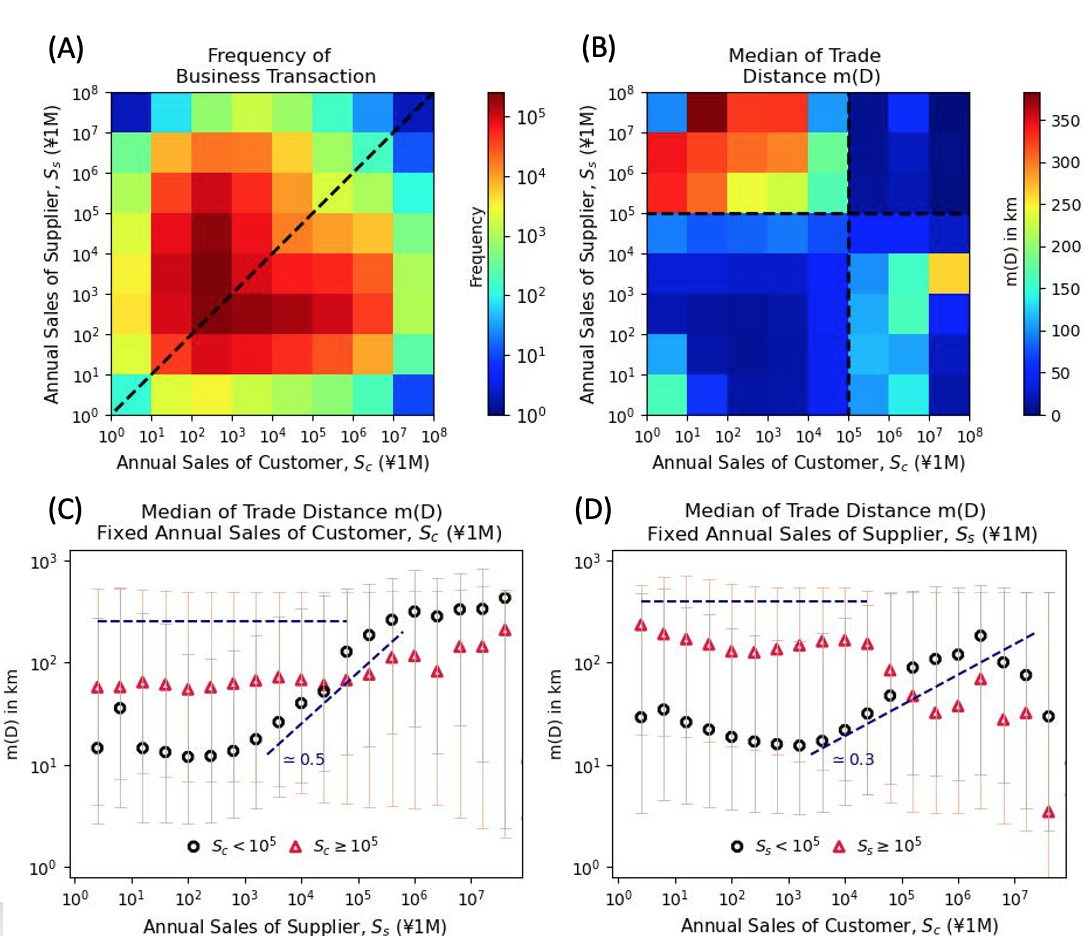} 
    
    \caption{ {\bf The frequency of trade and distances $D$ highly depend on the annual sales of customers $S_{c}$ and annual sales of suppliers $S_{s}$ for the year 2010, and similarly to 2021.} \
      ({\bf A}) is a heatmap showing the frequency trades among the $S_{c}$ and $S_{s}$ pairings, where the dotted diagonal line is the symmetry $S_{c}$ = $S_{s}$.
      ({\bf B}) is the equivalent heatmap quantifying the the median $m(D)$ of the $S_{c}$ and $S_{s}$ pairing, where the horizontal and vertical lines indicate the break of the $S_{c} \simeq S_{s}$ symmetry.
      ({\bf C}) shows the scaling of the median distance $m(D)$ of firms at different $S_{s}$ levels (above and below the symmetry break) with fixed $S_{c}$, and\
      ({\bf D}) is an equivalent plot with fixed $S_{s}$ and variable $S_{c}$ levels. \
      Each black circle and red triangle represents firms where $S < 10^{5}$ and $S \geq 10^{5}$, respectively.\
      Figures are plotted on a log-log scale for the 2010.\
  \label{fig:2s}       }
  \end{center}   
\end{figure}

Fig.\ref{fig:2s} represents the year 2010 and it was generated on an exact basis to Fig. 1 within the research paper (the year 2021). We include in this Supplementary Information to show that both figures are largely similar and that the break in symmetry and other trends noted in our analysis remain valid regardless of the year in question.

\section*{S3: Correlation between the exponents of sizes of Economic Zones, ⟨R⟩, and the decay $\gamma$ in the normalised probability distributions $P_{t}/P_{r}$}
\label{sub:ld}

\begin{figure}[b!]
  \begin{center}
    \includegraphics[width=0.8\hsize]{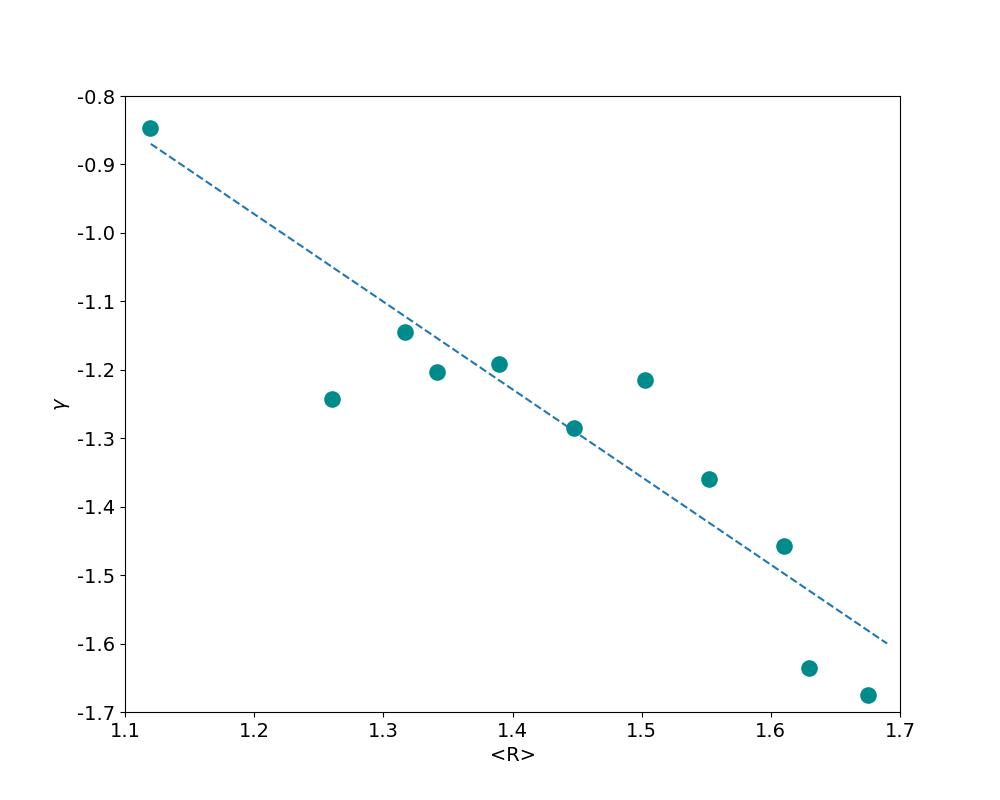} 
    
    \caption{ {\bf Correlation between the size of economic zones $⟨R⟩$ and power law exponent $\gamma$.} \
      The figure was produced by binning the x-axis into a fixed number of datapoints (5), and calculating the averages of $⟨R⟩$ and $\gamma$ within each bin.\
      The dotted line represents the best fit line, excluding the outlier point around (1.5 , -1.12).\
  \label{fig:5s}       }
  \end{center}   
\end{figure}

Table t1 shows the exponent of the size of economic zones $⟨R⟩$ and power law exponent $\gamma$ observed in the normalised probability distributions of trade distance $P_{t}/P_{r}$ for each prefecture in 2021. As described in our research paper, we observed a significant level of correlation between the pairing (⟨R⟩,$\gamma$), which is illustrated by Fig. \ref{fig:5s} (with one outlier point). 

\begin{figure}[b!]
  \begin{center}
    \includegraphics[width=1\hsize]{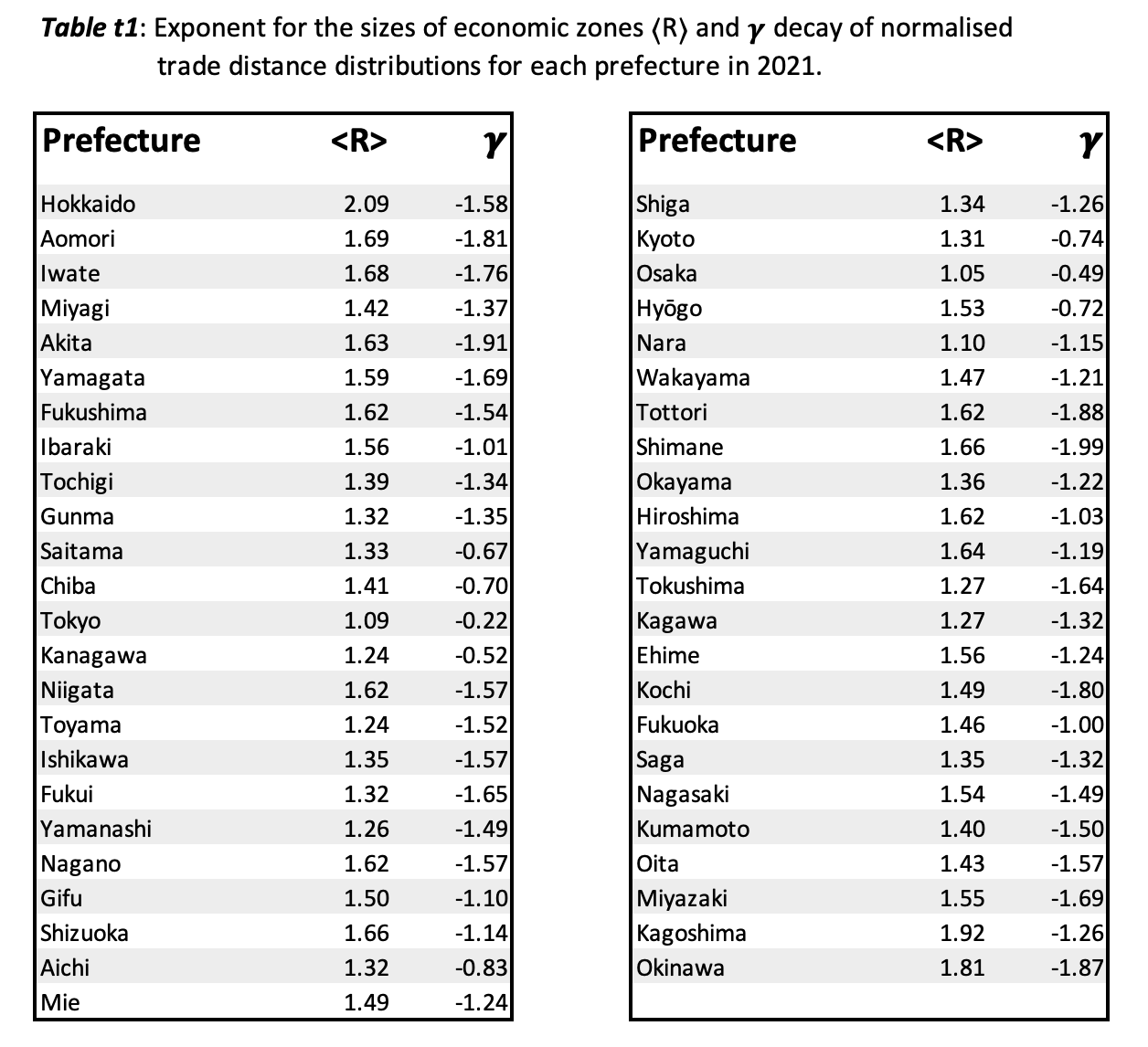} 
    
  \end{center}   
\end{figure}

\clearpage

\section*{S4: Distribution of companies across prefectures in Japan}

\begin{figure}[b!]
  \begin{center}
    \includegraphics[width=1\hsize]{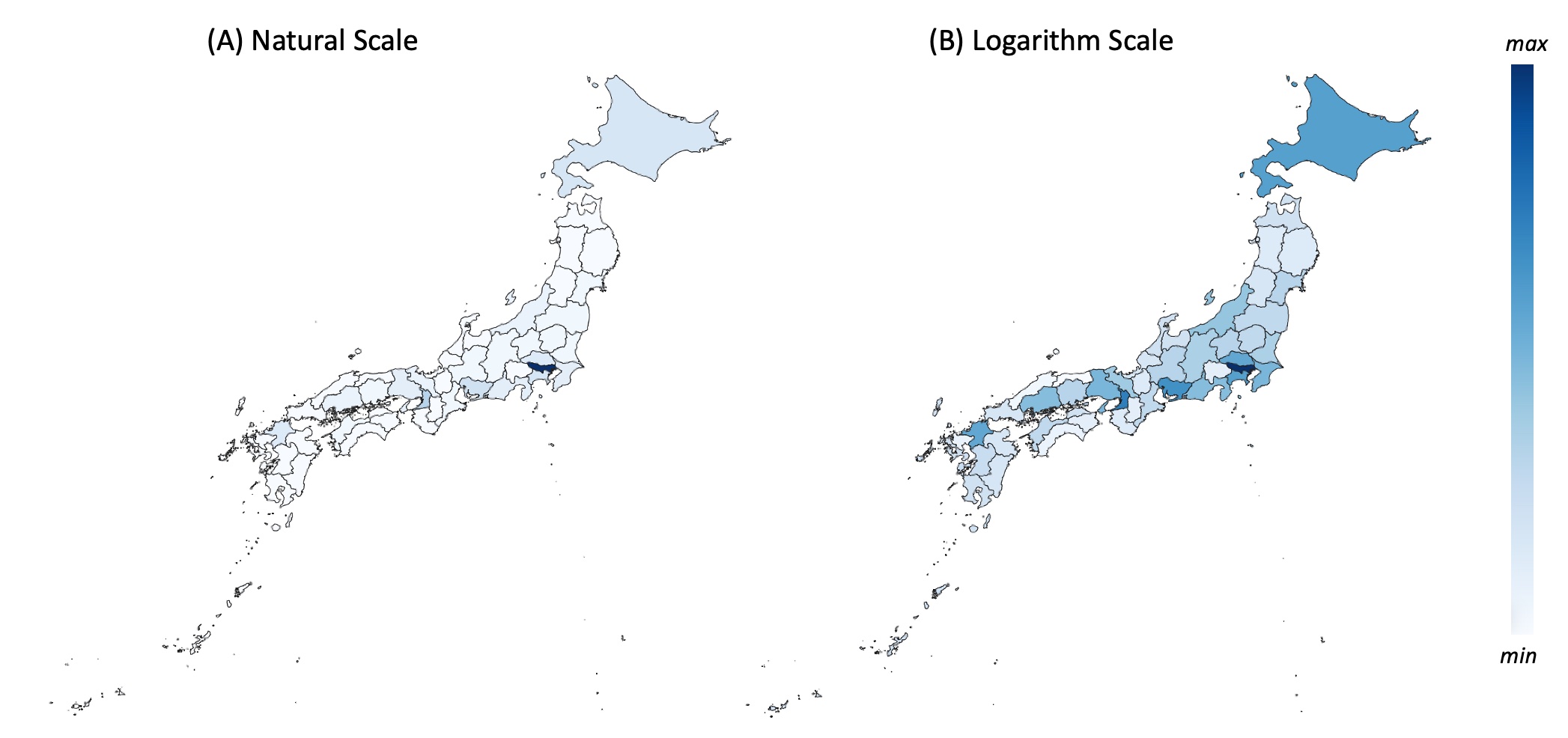} 
    
    \caption{ {\bf Distribution of companies by prefecture in Japan 2021.} \
      ({\bf A}) shows the relative distribution of companies in Japan coloured in natural numbers. In contrast, 
      ({\bf B}) show the same distribution but colour-coded in $log_{10}$ scale.\
  \label{fig:3s}       }
  \end{center}   
\end{figure}

\begin{figure}[b!]
  \begin{center}
    \includegraphics[width=1\hsize]{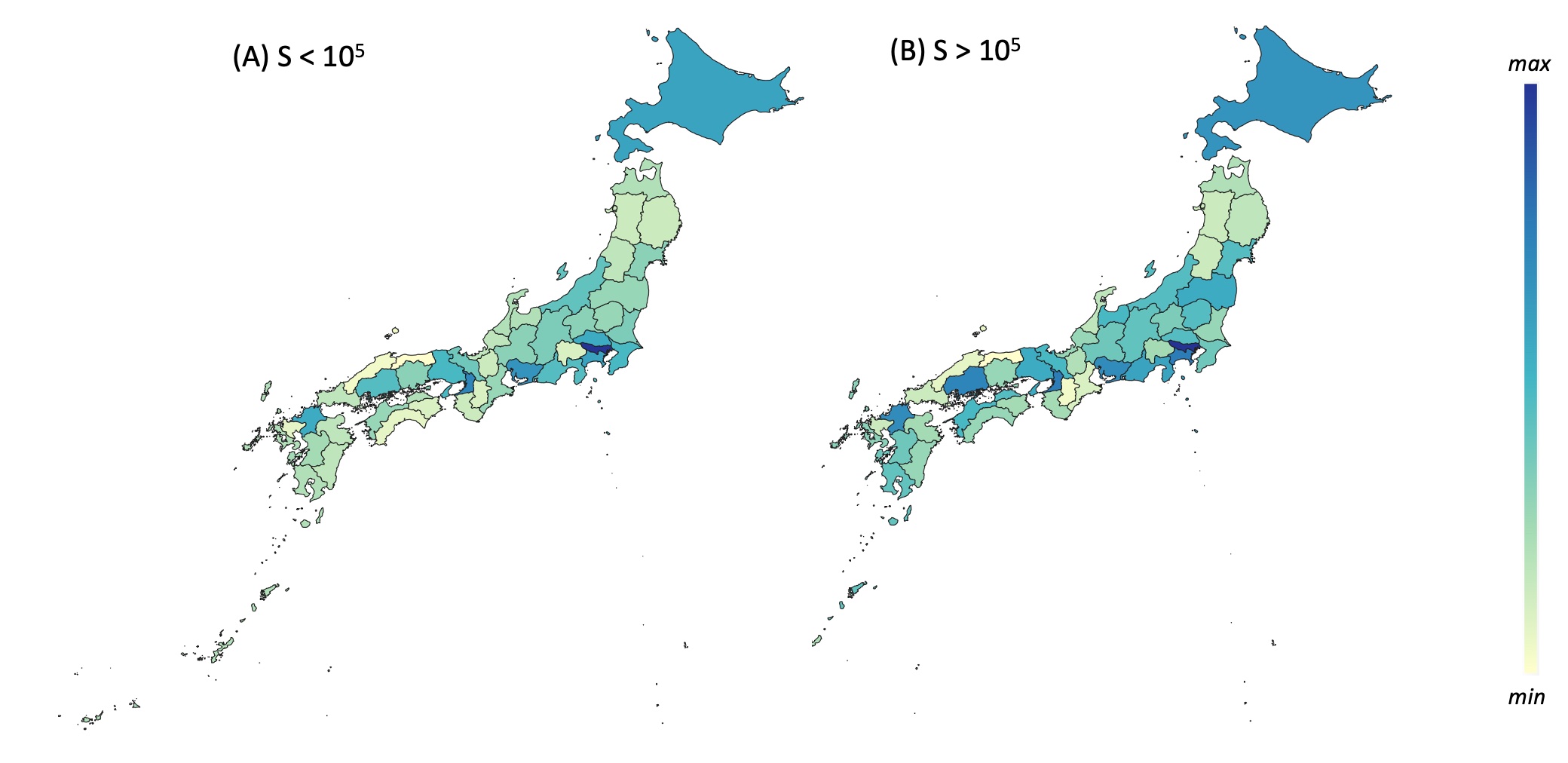} 
    
    \caption{ {\bf Distribution of companies by prefecture in Japan 2021, split by the annual sales, $S$.} \
      ({\bf A}) shows the distribution of companies in Japan on a $log_{10}$ scale for companies with annual sales below $10_{5}$ yen, whereas 
      ({\bf B}) is the equivalent map for the companies with annual sale above $10_{5}$ yen.\
  \label{fig:4s}       }
  \end{center}   
\end{figure}

The maps within Figs. \ref{fig:3s} and \ref{fig:4s} are solely an illustrative reference to the reader with regards to basic company statistics in Japan. It is worth to point our however (a) the large concentration within the Tokyo prefecture, as described in our research paper; and (b) the fact that there is no major obvious differences in the distribution  of smaller and middle size firms when compared to that of the largest firms.

